\documentclass{aa}
\usepackage{epsf}
\usepackage{graphicx}
\usepackage{txfonts}
\usepackage{natbib}
\usepackage{longtable}
\usepackage{xspace}
\bibpunct{(}{)}{;}{a}{}{,}   

\newcommand{\kms}{km~s$^{-1}$}

\newcommand{\teff}{\ensuremath{T_{\mathrm{eff}}}}

\defcitealias{PHM2004}{PHM}

\begin{document}
\title{On near Chandrasekhar mass central stars of planetary nebulae}

\author{R.~Napiwotzki}
\offprints{R.~Napiwotzki (e-mail: rn@star.herts.ac.uk)}
\institute{Centre for Astrophysics Research, STRI, University of Hertfordshire,
College Lane, Hatfield AL10 9AB, UK}

\date{Received,  accepted }

\abstract
{}
{A recent spectroscopic analysis of central stars of planetary nebulae
  (CSPNe) claims that the sample includes five CSPNe with masses very
  close to the Chandrasekhar limit of white dwarfs.
This claim should be verified or discarded from the available
  kinematical and chemical abundance information.}
{Kinematical parameters are extracted from Galactic orbits and
  compared with parameters expected for populations of different
  ages. The chemistry of the nebulae is compared with average values for
  different types.}
{The reported high masses are not supported by our investigation. The claimed
high central star masses are in contradiction with all other
evidence. A more consistent picture emerges if CSPN masses close
to the peak of the white dwarf mass distribution are assumed.}
{}


\maketitle

\section{Introduction\label{s:intro}} 
 
Stellar masses are the most fundamental parameter of stars. 
However, they are notoriously difficult to derive by direct methods.
The large majority of mass determinations make
use of stellar models derived from computations of their
evolution. Therefore it is of utmost importance that these models are
checked against independent mass determinations.

In this respect the situation for planetary nebulae (PNe) and their
central stars (CSPNe) is mostly dire. \citet{Nap1999} performed an
NLTE model atmosphere analysis of high gravity CSPNe and determined
masses from the post-AGB tracks of \citet{Blo1995b} and
\citet{Sch1983}. In a subsequent study \cite{Nap2001} showed that the
resulting distances are consistent with trigonometric parallax
measurements and distances determined from companions in wide
binaries, thus confirming the parameter and mass
determination. However, this comparison was limited to high gravity
central stars, which had already entered the white dwarf cooling
sequence. No such test has been performed for low gravity CSPNe on the
constant luminosity part of the post-AGB evolution.  ``Statistical''
distance determinations using properties of the PN, like the famous
Shklovsky method, exist, but they are notoriously unreliable
\citep{Nap2001} and therefore cannot provide an independent test.

In a recent study \citet[ hereinafter PHM ]{PHM2004} performed a model
atmosphere analysis of UV spectra of nine luminous CSPNe. This
analysis is based on spherically symmetric NLTE model atmospheres,
which treat the hydrodynamics of the stellar wind in a self-consistent
way. These and similar model atmospheres have been successfully applied
to the spectra of hot massive main sequence stars \citep[see
  discussion in][]{PHM2004}. 
The beauty of this method comes from the fact that it enables the
determination of the absolute quantities stellar mass $M$ and radius
$R$ without reference to stellar structure models. $M$ and $R$ are
closely linked to the terminal wind velocity $v_\infty$ and the mass
loss rate $\dot{M}$, which can be determined from a fit of the spectral lines. 

\begin{table*}
\caption{Stellar parameters of the high mass CSPN from
  \citetalias{PHM2004}, heliocentric radial velocities 
  \citep[from ][]{DAZ1998} and
  proper motion measurements. The last column gives the source of the
  proper motion values.}
\label{t:params}
\begin{tabular}{llccclr@{$\pm$}lr@{$\pm$}lr@{$\pm$}ll}\hline\hline
PN\,G        &common name         &\teff  &$\log g$ 
                &$M$ &$d$ &\multicolumn{2}{c}{$v_\mathrm{rad}$}
    &\multicolumn{2}{c}{$\mu_\alpha$} &\multicolumn{2}{c}{$\mu_\delta$}
    &source\\
        &  &(K)      &($\mathrm{cm\,s^{-2}}$) &($M_\odot$) &(kpc) 
        &\multicolumn{2}{c}{($\mathrm{km\,s^{-1}}$)}
    &\multicolumn{2}{c}{($\mathrm{mas\,yr^{-1}}$)}
    &\multicolumn{2}{c}{($\mathrm{mas\,yr^{-1}}$)}\\ \hline
025.3+40.8   &IC\,4593         &40\,000  &3.80   &1.11  &3.63
             &22.0&0.5      &$-$8.7&1.6   &4.1&1.5 &TY2\\
083.5+12.7   &NGC\,6826        &44\,000  &3.90   &1.40  &3.18
             &$-$6.2&0.6    &$-$11.0&2.9  &$-$9.7&1.7 &TRC\\
215.2$-$24.2 &IC\,418          &39\,000  &3.70   &1.33  &2.00
             &61.9&0.5      &$-$1.2&1.7   &2.5&1.8 &TRC\\
315.1$-$13.0 &Hen\,2-131       &33\,000  &3.10   &1.39  &5.62
             &$-$1.2&4.4    &$-$2.1&1.6   &$-$5.1&1.6 &TY2\\
316.1+08.4   &Hen\,2-108       &39\,000  &3.70   &1.33  &6.76
             &$-$11.1&0.4   &$-$0.6&2.7     &$-$1.1&2.5 &TY2\\
345.2$-$08.8 &IC\,1266, Tc\,1  &35\,000  &3.62   &1.37  &3.73
             &$-$84.1&4.7   &$-$1.6&2.2     &$-$11.4&2.2  &TY2\\ \hline
\end{tabular}\\
TY2: Tycho-2 catalogue \citep{HFM2000}; TRC: Tycho reference catalogue
\citep{HKB1998}
\end{table*}

The \citetalias{PHM2004} analysis of the CSPNe sample produced a quite
astonishing result (Table~\ref{t:params}): five out of the nine
analysed stars have a mass very close to the Chandrasekhar limit for
white dwarfs \citep[$1.40M_\odot$;][]{HS1961} and a sixth one has a
mass of $1.1M_\odot$.  This result is astonishing in at least two
ways: 1) white dwarfs of such a high mass are quite rare and thus we
wouldn't expect such a high fraction in a sample of CSPNe, and 2) the
masses derived by \citetalias{PHM2004} differ quite considerably from
what is derived from the comparison of effective temperature $\teff$
and gravity $g$ with theoretical post-AGB tracks. If the results of
\citetalias{PHM2004} can be proofed correct, this would indicate
substantial flaws in today's theory of CSPNe evolution.

Extraordinary claims need extraordinary proof. How can the
\citetalias{PHM2004} results be verified? One possible test is the
comparison of the derived spectroscopic distances with other
independent measurements. However, as already discussed in
\citetalias{PHM2004} there is a deplorable lack of any reliable
distance determinations for stars from the \citetalias{PHM2004} sample
in particular and for CSPNe of similar type in general.

In this article I will take a different pathway. The initial
mass-final mass (IMFM) relation for white dwarfs is well established
\citep{Wei2000}. Some uncertainty exists at the high mass end, but we
know from observations of white dwarfs in open clusters and binaries
that high mass white dwarfs are produced by high mass
progenitors. Even if one allows for some scatter in the IMFM relation
and blue straggler scenarios it is hard to imagine that the
progenitor of a $>1M_\odot$ CSPNe is older than 1\,Gyr, only slightly
shorter than the main sequence lifetime of a $2M_\odot$ star
\citep{SSM1992}. Therefore the high mass CSPNe should obviously belong
to the young thin disk population of the Milky Way. This population is
characterised by small peculiar velocities (ie.\ a small velocity
dispersion) and small scale heights perpendicular to the Galactic
disk. Progenitors of high mass CSPNe are associated with dredge-up of
elements produced by nuclear burning, which can be detected by an
analysis of the CSPNe or the surrounding PNe.

I will use the kinematical properties and the PN chemistry of the 
\citetalias{PHM2004} sample for a sanity check. In Sect.~\ref{s:kin}
the kinematics of this sample is compared to the expectation for a
young population. Sect.~\ref{s:abund} presents the evidence available
from nebular abundance measurements. Sect.~\ref{s:discussion} will
conclude with a discussion of the combined evidence.

\section{Kinematics\label{s:kin}}

The kinematical state of a sample of stars contains information on the
population membership and their age. The motion of stars in the Milky
Way is usually described in the orthogonal coordinate system $X, Y, Z$
and the corresponding velocities $U, V, W$. $X$ points from the Sun in
the direction of the Galactic centre, $Y$ points in the direction of
the galactic rotation at the position of the Sun and $Z$ towards the
north Galactic pole. Here I will assume 8\,kpc distance of the
Sun from the Galactic centre, an orbital velocity of the local
standard of rest (LSR) of 220\,\kms and $U_\odot=10.0$\,\kms,
$V_\odot=7.2$\,\kms, $W_\odot=7.2$\,\kms relative to the LSR.

Older populations are characterised by larger scale heights in $Z$
direction and larger velocity dispersions in all
directions. Investigations of solar neighbourhood stars make often use
of diagrams plotting various combinations of the velocities
$U,V,W$. However, this would produce misleading results, if applied to stars
several kpcs away from the Sun. 

\begin{figure}
\resizebox{0.92\hsize}{!}{\includegraphics[angle=-90,bb=40 16 560 784]{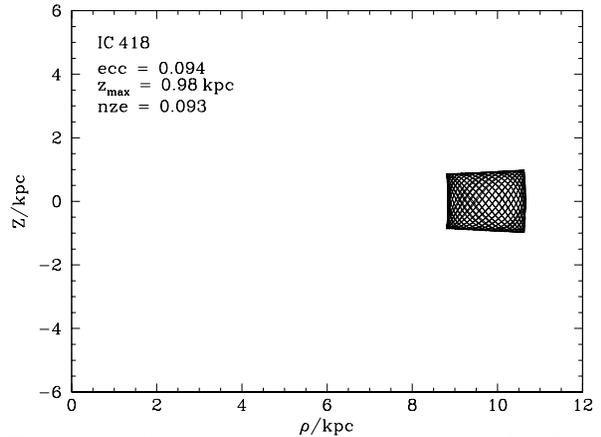}}
\caption{Meridional cut of the Galactic orbit of IC\,418. The orbit is 
integrated over 5\,Gyr.} 
\label{f:ic418}
\end{figure}
\begin{figure}
\sidecaption
\resizebox{0.92\hsize}{!}{\includegraphics[angle=-90,bb=40 16 560 784]{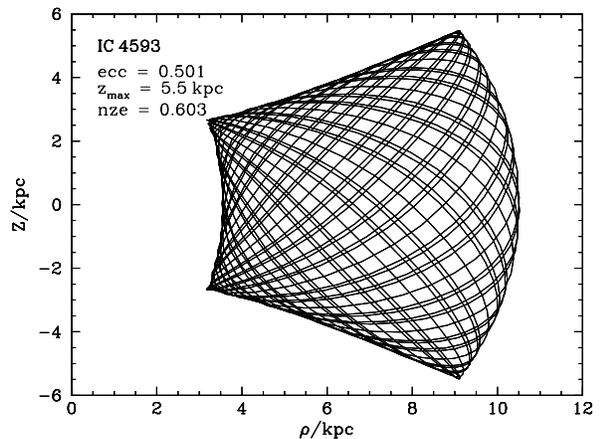}}
\caption{Meridional cut of the Galactic orbit of IC\,4593.}
\label{f:ic4593}
\end{figure}

Insights into the kinematical state of stars can be gained from their orbits
in the Milky Way. The orbits presented here are calculated with the programme
{\sc orbit6} \citep{OB1992}. The Galactic potential of \citet{AS1991} is
adopted. Radial velocities and proper motions were compiled from literature
and are presented in Table~\ref{t:params}. Distances were taken from
\citetalias{PHM2004}. The results for IC\,418 and IC\,4593 are shown in
Figs.~\ref{f:ic418} and~\ref{f:ic4593}, respectively. We use so-called
meridional cuts with $\rho = \sqrt{X^2+Y^2}$ as horizontal axis. Note
that the orbit of a thin disk star without any peculiar velocities would
appear as a dot in these plots. The orbit of IC\,418 (Fig.~\ref{f:ic418})
shows the characteristics of a disk star. The eccentricity $ecc$, defined as 
\begin{displaymath}
ecc = (R_{\mathrm{a}} - R_{\mathrm{p}})/
           (R_{\mathrm{a}} + R_{\mathrm{p}})
\end{displaymath}
with $R_{\mathrm{a}}$ and $R_{\mathrm{p}}$ being the apo- and
perigalactic distances, is significant, but not extreme. The maximum
distance from the Galactic plane $z_{\mathrm{max}}$ is on the high
side and makes membership in the old thin disk or the thick disk
possible.  The orbit of IC\,4593 (Fig.~\ref{f:ic4593}) is more extreme
reaching large distances from the Galactic plane and having high
eccentricity.  This points to a halo nature of this star. The
discussion of bigger samples and error limits is facilitated by
condensing the properties of the orbits into simple numbers. I will
make use of the eccentricity $ecc$ and normalised $z$-extent
\begin{displaymath}
nze = z_{\mathrm{max}}/ \rho(z_{\mathrm{max}})
\end{displaymath}
introduced by \citet{dBAA1997} to take the effect of the Galactic potential
diminishing with galactrocentric distance into account.

\begin{figure}
\resizebox{0.92\hsize}{!}{\includegraphics[angle=-90,bb=40 16 560 784]{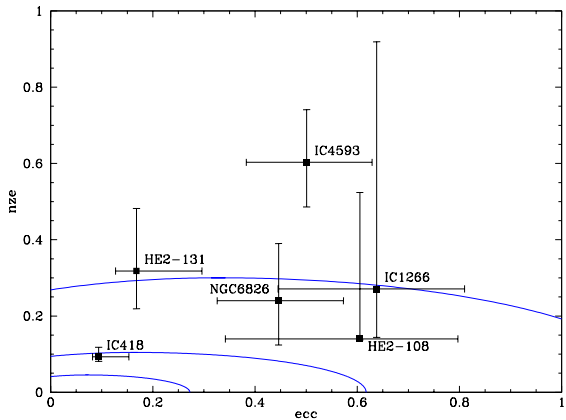}}
\caption{Kinematical properties of the CSPNe in the eccentricity-$nze$
diagram. The distances computed by \citetalias{PHM2004} and measured
proper motions from Table~\ref{t:params} were used for this plot. The contour
lines show the $3\sigma$ limits for young disk, old disk and thick disk stars
(in order of increasing size). The error bars for individual objects 
indicate 68\% confidence intervals.}
\label{f:kinobs}
\end{figure}

Values of $ecc$ and $nze$ for the CSPNe sample are plotted in
Fig.~\ref{f:kinobs}. The measurement errors were propagated via a Monte
Carlo simulation (distance errors of 15\% were assumed). For a
discussion of these results we need to know where to find populations
of different age in this diagram. We started with the velocity
dispersions and asymmetric drifts given by \citet{RRD2003} for the
young thin disk, the old thin disk and the thick disk
(Table~\ref{t:pop}) and translated this into distributions in $ecc$
and $nze$ via a Monte Carlo simulation with the {\sc orbit6}
programme. 


\begin{table}
\caption{Adopted velocity dispersions $\sigma_U$, $\sigma_V$,
  $\sigma_W$ and asymmetric drift $v_{\mathrm{ad}}$ for the young thin 
  disk (YTD), the old thin disk (OTD) and for the thick disk
  \citep[from ][]{RRD2003}.}\label{t:pop}
\begin{tabular}{lrrrr}\hline\hline
population  &$\sigma_U$ &$\sigma_V$
         &$\sigma_W$ 
         &$v_\mathrm{ad}$\\ 
         &$(\mathrm{km\,s^{-1}})$  &$(\mathrm{km\,s^{-1}})$
         &$(\mathrm{km\,s^{-1}})$  &$(\mathrm{km\,s^{-1}})$ \\ \hline
YTD (0.15-1Gyr)   &19.8 &12.8 &8.0 &3.1\\
OTD (7-10\,Gyr)   &43.1 &27.8 &17.5 &14.8\\
thick disk        &67   &51   &42   &53\\ \hline
\end{tabular}
\end{table}

The results are shown in Fig.~\ref{f:kinobs}. They are completely
contrary to what we expect for a young sample. Only one CSPNe
(IC\,418) is marginally consistent with a membership in the old thin
disk, Fig.~\ref{f:kinobs} indicates even a halo nature of the
supposedly young CSPNe of IC\,4593! Admittedly, the error bars are
large. The largest contribution comes from the uncertainties of the
proper motion measurement. In principle the measurement of proper
motions on this level of accuracy can be hampered by nebular structure
close to the central star, which could cause offsets. If one wants to
play devil's advocate one could argue that the proper motions are
completely unreliable. Although this is very likely over-pessimistic,
it is possible that the error bars underestimate the real errors.

\begin{figure}
\resizebox{0.92\hsize}{!}{\includegraphics[angle=-90,bb=40 16 560 784]{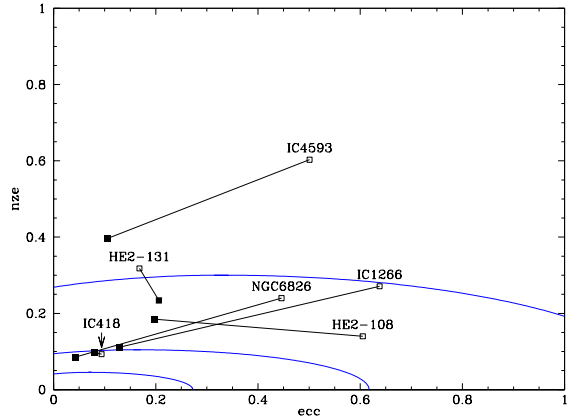}}
\caption{Kinematical properties of the CSPNe in the eccentricity-$nze$
diagram with proper motions treated as free parameters (filled
squares). The parameters resulting from the observed proper motions
are indicated by open squares connected to the new results.}
\label{f:kinfree}
\end{figure}

Thus I performed an experiment by treating the proper motions as free
parameters. Distance and radial velocities were kept fixed at their
values from Table~\ref{t:params} and an optimisation was performed to
minimise the combined values of $nze$ and eccentricity, $\sqrt{ecc^2
+nze^2}$. The result is shown in Fig.~\ref{f:kinfree}. Even now only
three CSPNe can be found close to the old thin disk contour, not to
mention the young thin disk. Three stars (He\,2-108, He\,2-131, IC\,4593)
must be members of the thick disk or halo populations with
a very high level of confidence. Thus we conclude that the kinematical
evidence is in clear contradiction with the CSPNe being part of a
young population.

\section{Nebular abundances\label{s:abund}}

High mass central stars with high mass progenitors are usually
identified with PNe of type~I in the classification scheme of
\citet{Pei1978}. These show enhanced helium and nitrogen abundances 
\citep[$\mathrm{He/H} >0.125$ and $\mathrm{N/O} >0.50$;][]{PT1983},
which are explained by dredge-up processes in the progenitor. Although
it is by no means clear, whether {\em all} PN with high mass CSPN have
enhanced abundances of helium and nitrogen, evidence for a type~I
nature of the \citetalias{PHM2004} objects would strongly support
their high mass nature. 

\begin{table}
\caption{Nebular abundances for the very high mass CSPN of the
  \citetalias{PHM2004} sample.}
\label{t:abund}
\begin{tabular}{lrlc}\hline\hline
PN           &He/H     &N/O     &$\log(\mathrm{O/H})+12$ \\ \hline
IC\,4593     &0.099    &0.047   &8.50\\
NGC\,6826    &0.107    &0.10    &8.46\\
IC\,418      &0.086    &0.13    &8.76\\
Hen\,2-131   &         &0.38    &8.67\\
Hen\,2-108$^\mathrm{a}$ &$>0.117$   &0.28    &8.40\\
IC\,1266$^\mathrm{b}$     &0.074    &0.09    &8.71\\ \hline
type I       &0.137    &0.85    &8.66\\
type II-III  &0.103    &0.26    &8.66\\ \hline
\end{tabular}\\
$^\mathrm{a}$ \citet{TP1977}; $^\mathrm{b}$ \citet{Phi2003}
\end{table}

Chemical abundances determined from the analysis of the surrounding
nebulae are compiled in Table~\ref{t:abund}. These were taken from the
compilation of \citet{Per1991} unless noted
otherwise. Table~\ref{t:abund} is supplemented by average abundances
of type~I and type~II-III PNe from \citet{Per1991}. PNe of type~II
and~III are identified with lower mass CSPNe from the Galactic disc,
probably including some objects from the thick disk as well. 

Table~\ref{t:abund} reveals that none of the PNe of the
\citetalias{PHM2004} sample qualifies as type~I. The chemical abundances are
well within the range expected for run of the mill PNe of type~II
and~III. We conclude that again a high mass nature of the investigated
objects is not supported. 

\section{Putting it all together: What is the true nature of these
  objects?\label{s:discussion}}

This article presented two tests to verify or disprove the results of
\citetalias{PHM2004} that six of the CSPNe analysed by them have very
high masses. The kinematical results of Sect.~\ref{s:kin} are clearly
at odds with these CSPN being young objects with massive progenitor
stars. The PNe chemistry presented in Sect.~\ref{s:abund} also points
to lower mass progenitors of the \citetalias{PHM2004} CSPNe. Viewed
together I conclude that the \citetalias{PHM2004} sample has failed
this test and the high masses are not confirmed.

\begin{figure}
\resizebox{0.92\hsize}{!}{\includegraphics[angle=-90,bb=40 16 560 784]{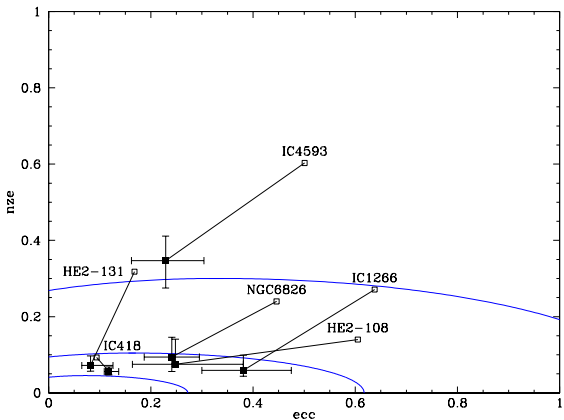}}
\caption{Kinematical properties of the CSPNe in the eccentricity-$nze$
diagram resulting for assumed CSPNe masses of $0.565M_\odot$ 
as discussed in the text.}
\label{f:kinshort}
\end{figure}

Now, are all these stars members of the old and metal poor thick disk
and halo populations? This seems unlikely, given that the PN abundance
determinations in Table~\ref{t:abund} show that these are not very
metal poor objects.  However, the halo-like kinematics of the
\citetalias{PHM2004} sample are largely the result of the very large
distances computed from the \citetalias{PHM2004} parameters
(Table~\ref{t:params}). The picture becomes very different, when we
assume a low mass close to the peak of the white dwarf mass
distribution \citep[$0.56\ldots0.59M_\odot$][]{NGS1999,LBH2005} and a
radius consistent with standard stellar structure calculations. To
demonstrate the effect I computed distances resulting from the
$0.565M_\odot$ post-AGB track of \citet{Sch1983} and repeated the
orbit calculations with these values. Note that I could have adopted
the parameters of \citet{KMP1997} instead, but it is not impossible that
these analyses have there own caveats, as discussed by the authors.
The result is shown in
Fig.~\ref{f:kinshort}. Now the parameters of five CSPNe are consistent
with being members of the old thin disk. IC\,4593 maybe member of the
thick disk or halo. Interestingly, the oxygen abundances of this PN is
somewhat lower than normal (Table~\ref{t:abund}), which would be
consistent with this finding. On the other hand the kinematical
properties of single objects are no firm proof of their
affiliation. They could have, for example, suffered from particular 
violent encounters with other stars.

We conclude that neither the kinematical evidence nor the PNe
chemistry supports the claim of high masses of the six CSPNe from the
\citetalias{PHM2004} sample. The findings are far better explained, if
one assumes much lower CSPNe masses, which would correspond to lower
progenitor masses. Unfortunately this would indicate that the model
atmospheres applied by \citetalias{PHM2004} are not yet able to produce
reliable results for CSPNe, in contrast to their successful
application to massive stars.
\acknowledgement 
{R.N.\ gratefully acknowledges support by a PPARC Advanced Fellowship.}

\bibliographystyle{aa}
\bibliography{star,own}

\end{document}